\documentclass{ws-ijmpe}
\begin{document}
\markboth{D. Singh, G. Saxena, M. Kaushik, H. L. Yadav and H. Toki}
{Study of Two-proton Radioactivity Within the RMF+BCS Approach}
\catchline{}{}{}{}{}
\title{STUDY OF TWO-PROTON RADIOACTIVITY WITHIN THE RELATIVISTIC MEAN-FIELD PLUS BCS APPROACH}
\author{D. SINGH}
\address{Department of Physics, University of Rajasthan,
Jaipur-302004, India\\ dsingh5@gmail.com}
\author{G. SAXENA}
\address{Department of Physics, Govt. Women Engineering College,
Ajmer-305002, India}
\author{M. KAUSHIK}
\address{Department of Physics, University of Rajasthan,
Jaipur-302004, India}
\author{H. L. YADAV}
\address{Department of Physics, Banaras Hindu University,
Varanasi-221005, India\\ hlyadavphysics@gmail.com}
\author{H. TOKI}
\address{Research Center for Nuclear Physics (RCNP), Osaka University, 10-1,
Mihogaoka, Ibaraki, Osaka 567-0047, Japan}

\maketitle
\begin{history}
\received{Day Month Year} \revised{Day Month Year}
\end{history}

\begin{abstract}
Inspired by recent experimental studies of two-proton radioactivity
in the light-medium mass region, we have employed relativistic
mean-field plus state dependent BCS approach (RMF+BCS) to study the
ground state properties of selected even-Z nuclei in the region 20 $
\leq$ Z $\leq$ 40. It is found that the effective potential barrier
provided by the Coulomb interaction and that due to centrifugal
force may cause a long delay in the decay of some of the nuclei even
with small negative proton separation energy. This may cause the
existence of proton rich nuclei beyond the proton drip-line. Nuclei
$^{38}$Ti, $^{42}$Cr, $^{45}$Fe, $^{48}$Ni, $^{55}$Zn, $^{60}$Ge,
$^{63,64}$Se, $^{68}$Kr, $^{72}$Sr and $^{76}$Zr are found to be the
potential candidates for exhibiting two-proton radioactivity in the
region 20 $ \leq$ Z $\leq$ 40. The reliability of these predictions
is further strengthened by the agreement of the calculated results
for the ground state properties such as binding energy, one- and
two-proton separation energy, proton and neutron radii, and
deformation with the available experimental data for the entire
chain of the isotopes of the nuclei in the region 20 $ \leq$ Z
$\leq$ 40.

\keywords{Relativistic mean-field theory; Two-proton radioactivity;
One- and Two-proton separation energy; Proton drip-lines.}
\end{abstract}
\ccode{PACS Nos.: 23.50.+z, 21.10.-k, 21.10.Dr}

%
%

\section{Introduction}
\label{sect:intro}
The structure and decay modes of nuclei at and beyond the proton
drip-line represent one of the most active areas in both
experimental and theoretical studies of exotic nuclei with extreme
isospin values. At the proton drip-line, further addition of protons
is not possible as the nucleus becomes unbound. Beyond the
drip-line, one or more valence protons may still remain confined due
to Coulomb and centrifugal barriers enabling the nucleus to acquire
rather long mean life time. Subsequently, it may decay by the
process wherein one or more protons tunnel through the barrier
leading to observation of one or more protons radioactivity. This
situation is quite different from the neutron rich side of the
valley of $\beta$-stability where Coulomb barrier is absent and
consequently the drip-line gets extended to highly neutron rich
nuclei.

Decay modes  of nuclei through one- or two- proton radioactivity
were theoretically proposed in the early 1960's for the first time
by Goldansky \cite{goldansky}. The one-proton radioactivity
predicted for odd-proton nuclei was indeed observed already in the
early 1980's in experiments carried out at GSI, Darmstadt
\cite{hofmann}, and presently many nuclei  (more than 30) which
decay in their ground state by one-proton emission are well known
\cite{Blank1}. However, the two-proton emission  mode was
experimentally verified only  more than four decades after its
theoretical prediction in the decay of $^{45}$Fe
\cite{Pftzner,Giovinazzo}, and subsequently in other experiments in
the decay of $^{54}$Zn \cite{Blank,Ascher} and $^{48}$Ni
\cite{Dossat,Pomorski}. Many more experiments are being carried out
to discover the new candidates.

Two-proton radioactivity occurs when the sequential emission of two
independent protons from the nuclear ground state is energetically
forbidden, but the emission of a pair of protons is allowed. In this
situation, due to the gain of stability from the pairing energy, the
mass of the even-Z two-proton emitter is smaller than the mass of
the odd-Z one-proton daughter giving rise to the negative Q value
for one-proton emission. Such a situation prohibits one-proton
emission and favors two-proton radioactivity. Therefore, the
simultaneous two-proton emission is energetically possible only
beyond the two-proton drip-line and competes with two other decay
modes, namely one-proton emission and $\beta$-decay. During the
emission process, the two protons must tunnel through a wide Coulomb
barrier. Hence, the emission probability strongly depends on the
available Q$_{2p}$ value. For small Q$_{2p}$ values, $\beta$-decay
is more probable, whereas for the large ones the life time of the
given nucleus is very short leading to a fairly small window of
opportunity to observe the two-proton radioactivity. Therefore, one
has to look for nuclei in which one-proton decay is energetically
impossible. Consequently we have chosen to study the even-Z nuclei
only. Even-Z nuclei satisfying the condition S$_{p}$ $>$ 0 and
S$_{2p}$ $<$ 0 may be the possible candidates for simultaneous
two-proton emission.

Theoretical studies of one and two-proton radioactivity have been
carried out within the framework of different models
\cite{Blank1,mag1,mag2,lsf1,mag3,lsf2,bab1,orm,bar1,bar2,brw,grig1,grig2}.
Also, the relativistic Hartree-bogoliubov and relativistic mean
field (RMF) approaches \cite{lala1,lala2,lala3,lala4,geng0} have
been employed with reasonable success for the description and
prediction of one-proton radioactivity in the proton rich exotic
nuclei in the vicinity of the proton drip-line. However, such a
relativistic approach has not been used so far to describe and
predict the two-proton radioactivity in the proton rich exotic
nuclei as observed recently
\cite{Pftzner,Giovinazzo,Blank,Ascher,Dossat,Pomorski}. This has
motivated us to study the two-proton emitter nuclei within the
framework of RMF+BCS approach.
The main advantage of the RMF+BCS approach is that it provides the
spin-orbit interaction in the entire mass region in a natural way
\cite{walecka,boguta,serot,pgr}. This indeed has proved to be very
crucial for the study of unstable nuclei near the drip-line, since
the single particle properties near the threshold are prone to large
changes as compared to the case of deeply bound levels in the
nuclear potential. In addition to this, the pairing properties are
equally important for nuclei near the drip-line. As nuclei move away
from stability and approach the drip-lines, the corresponding Fermi
surface gets closer to zero energy at the continuum threshold. A
significant number of the available single-particle states then form
part of the continuum. Indeed the RMF+BCS scheme
\cite{yadav,yadav1,yadav2,yadav3} yields results which are in close
agreement with the experimental data and with those of continuum
relativistic Hartree-Bogoliubov (RCHB) and other similar mean-field
calculations \cite{meng,meng2}.

We have restricted our investigations to the proton rich isotopic
chains of even Z  nuclei in the medium mass region 20 $ \leq$ Z
$\leq$ 40 as most of the available experimental data belong to this
region \cite{Pftzner,Giovinazzo,Blank,Ascher,Dossat,Pomorski}.


\section{Theoretical Formulation and Model}
The relativistic mean-field (RMF) approach \cite{walecka,pgr,ring}
provides a description of the nuclear many-body system in terms of
an effective Lagrangian containing mesonic and nucleonic degrees of
freedom \cite{serot}. The relativistic mean-field model of the
nucleus is formulated on the basis of two approximations
\cite{serot,pgr2}: (i) the mean-field assumption and (ii) the no-sea
approximation.

\subsection{Model Lagrangian Density}

For our RMF calculations we have included apart from the photonic
field, the fields corresponding to the $\sigma,\, \omega, $ and
$\rho$ mesons \cite{walecka,pgr,ring} as shown in Table 1.

\begin{table}[h]
\tbl{Quantum numbers for different mesons used in the present
investigations.} {\begin{tabular}{|cccc|} \hline
\multicolumn{1}{|c}{Name of meson used in the Model}&
\multicolumn{1}{c}{Angular momentum($J$)}&
\multicolumn{1}{c}{Isospin($T$)}&
\multicolumn{1}{c|}{Parity($P$)}\\
\hline
$\sigma$&0&0&1 \\
$\omega$&1&0&-1 \\
$\rho$&1&1&-1 \\
\hline
\end{tabular}}
\end{table}

The Lagrangian density is written as the sum of the Lagrangian
density for free nucleons ${\cal L}^{free}_N$, the Lagrangian
density ${\cal L}^{free}_M$ for the free mesons ${\sigma}$,
${\omega}$ and ${\rho}$, and that of photon, and the Lagrangian
density for the interaction between mesons and nucleons ${\cal
L}_{int}$,
\begin{eqnarray}
   {\cal L}& = & {{\cal L}^{free}_N} + {\cal L}^{free}_M + {\cal L}_{int}
   \nonumber
\end{eqnarray}

The interaction term includes both the linear and nonlinear
couplings. For  the description of nonlinear couplings we follow the
treatment of Boguta and Bodmer \cite{boguta} in the case of
${\sigma}$-mesons, and that of Sugahara and Toki \cite{suga,suga2}
for the ${\omega}$-mesons. These nonlinear interaction terms have
been shown to play important role in various applications of the RMF
theory to provide a quantitative description of nuclear matter and
the ground state properties of nuclei
\cite{boguta,ring,pgr2,suga,suga2}. As mentioned earlier this is an
effective Lagrangian density to be used together with the mean-field
and no-sea approximations. Thus when written in detail the model
Lagrangian used in our present study has the following form,
\begin{eqnarray}
   {\cal L}& = &{\bar\Psi} [\imath \gamma^{\mu}\partial_{\mu}
                  - M]\Psi\nonumber\\
   &&+ \frac{1}{2}\, \partial_{\mu}\sigma\partial^{\mu}\sigma
   - \frac{1}{2}m_{\sigma}^2 \sigma^2  - \frac{1}{3}g_{2}\sigma
    ^3 - \frac{1}{4}g_{3}\sigma^4 -g_{\sigma}
    {\bar\Psi}  \sigma {\Psi}\nonumber\\
     &&-\frac{1}{4}H_{\mu \nu}H^{\mu \nu} + \frac{1}{2}m_{\omega}
      ^2 \omega_{\mu}\omega^{\mu} + \frac{1}{4} c_{3}
      (\omega_{\mu} \omega^{\mu})^2
   - g_{\omega}{\bar\Psi} \gamma_{\mu}\Psi
  \omega^{\mu}\nonumber\\
       &&-\frac{1}{4}G_{\mu \nu}^{a}G^{a\mu \nu}
   + \frac{1}{2}m_{\rho}
         ^2 R_{\mu}^{a}R^{{\mu}a}
   - g_{\rho}{\bar\Psi} \gamma_{\mu}\tau^{a}\Psi
          R^{\mu a}\nonumber\nonumber\\
       &&-\frac{1}{4}F_{\mu \nu}F^{\mu \nu}
   - e{\bar\Psi} \gamma_{\mu} \frac{(1-\tau_{3})}
       {2} A^{\mu} \Psi\,\,\nonumber\
       \label{t001}
\end{eqnarray}
where we have used throughout ${\hbar} = c = 1$. Here the field
tensors, H, G and F for the vector fields due to $\omega$, $\rho$
and photon are defined through
\begin{eqnarray}
      H_{\mu \nu} &=& \partial_{\mu} \omega_{\nu} -
                       \partial_{\nu} \omega_{\mu}\nonumber\\
      G_{\mu \nu}^{a} &=& \partial_{\mu} R_{\nu}^{a} -
                       \partial_{\nu} R_{\mu}^{a}
       -g_{\rho}\,\epsilon^{abc}R_{\mu}^{b}
                    R_{\nu}^{c} \nonumber\\
       F_{\mu \nu} &=& \partial_{\mu} A_{\nu} -
                       \partial_{\nu} A_{\mu}\,\,\nonumber\\
\end{eqnarray}

Furthermore, the symbols $M$, $m_{\sigma}$, $m_{\omega}$, and
$m_{\rho}$, are the masses of nucleon, and that of the $\sigma$,
$\omega$, and $\rho$ mesons, respectively. The superscript `a'
labels the isospin degree of freedom and runs from 1 to 3.
Similarly, $g_{\sigma}$, $g_{\omega}$, $g_{\rho}$ and $e^2/4\pi$ =
$1/137$ are the coupling constants for the mesons, and the photon,
respectively, whereas $\tau^a$ are the Pauli isospin matrices.

The set of parameters appearing in the effective Lagrangian include
(i) the masses of the nucleons and the mesons $M$, $m_{\sigma}$,
$m_{\omega}$ and $m_{\rho}$, (ii) the coupling constants of the
meson fields to the nucleons $g_{\sigma}$, $g_{\omega}$ and
$g_{\rho}$, and (iii) the parameters $g_2$ and $g_3$ which describe
the nonlinear coupling of the $\sigma$ mesons among themselves, and
the parameter $c_3$ which describes the nonlinear self coupling of
the vector meson $\omega$. These have been obtained in an extensive
study \cite{suga,suga2} which provides a reasonably good description
for the ground state of nuclei and that of nuclear matter
properties. This set, termed as TMA parameters, has an A-dependence
and covers the light as well as medium heavy nuclei from $^{16}O$ to
$^{208}Pb$. The TMA force parameter set determined in Ref.
\cite{suga,suga2} (displayed in Table 2) has been used in the
present RMF+BCS calculations.

\begin{table}[h]
\tbl{TMA force parameters along with the nuclear matter properties.}
{\begin{tabular}{|ccc|} \hline \multicolumn{1}{|c}{Parameters}&
\multicolumn{1}{c}{Unit}&
\multicolumn{1}{c|}{$TMA$}\\
\hline
M &(MeV)&938.9  \\
m$_{\sigma}$ &(MeV) &519.151\\
m$_{\omega}$ &(MeV) &781.950\\
m$_{\rho}$ &(MeV) &768.100\\
g$_{\sigma}$& & 10.055 + 3.050/A$^{0.4}$\\
g$_{\omega}$& & 12.842 + 3.191/A$^{0.4}$\\
g$_{\rho}$ & &  3.800 + 4.644/A$^{0.4}$\\
g$_{2}$ &(fm)$^{-1}$ & -0.328 - 27.879/A$^{0.4}$\\
g$_{3}$& & 38.862 - 184.191/A$^{0.4}$\\
c$_{3}$& &151.590 - 378.004/A$^{0.4}$ \\
\hline Nuclear Matter Properties &{}&{}\\
\hline

Saturation Density $\rho_{0}$&(fm)$^{-3}$ &0.147\\
Bulk binding energy/nucleon (E/A)$_{\infty}$&(MeV)&16.0\\
Incompressibility K  &(MeV) &318.0 \\
Bulk symmetry energy/nucleon a$_{Sym}$&(MeV)&30.68 \\
Effective mass ratio m$^{*}$/m & &0.635\\
\hline
\end{tabular}}
\end{table}

In the literature \cite{serot,ring,pgr2} many sets of
parameterizations are available for the Lagrangian density similar
to that of Eqn. (1) containing the nonlinear terms for the $\sigma$
mesons but without the inclusion of nonlinear potential for the
$\omega$ mesons. The earliest sets like NL1, NL2 and NL-Z etc. have
been employed extensively and further improved subsequently.

\subsection{Relativistic Mean-Field Equations}
The equations of motion for the RMF theory are obtained from the
variational principle by varying the action integral with respect to
the wave functions $\psi_i$ ($i = 1...A$) for the nucleons, and
fields $\sigma$, $\omega$ and $R$ for the mesons and the
electromagnetic field $A$. In general, for a given Lagrangian
density $\cal L$ the variational principle
\begin{eqnarray}
\delta \int{dt} \int d^{3}x\,{\cal L}(\vec{x} , t)&=& 0
\end{eqnarray}
leads for the variables $q_j$, the well known E.L. equations of
motion given by
\begin{eqnarray}
\frac{\partial{\cal L}}{\partial{q_j}} -
\partial_{\mu}\left\{\frac{\partial{\cal L}}{\partial(\partial_{\mu}{q_j})}\right\}
&=& 0
\end{eqnarray}
Using this we can derive the equations of motion for various fields
($q_j$ = $\psi_i,\, \sigma,\, \omega$, $R$ and $A$) appearing in our
effective Lagrangian density given by Eqn.~(\ref{t001}). Thus for
the nucleons fields $\psi_i$ one obtains the Dirac equation given by
\begin{eqnarray}
(\iota \gamma^{\mu}\partial_{\mu}- M - g_{\sigma}\sigma -
g_{\omega}\gamma^{\mu}\omega_{\mu}-g_{\rho}\gamma^{\mu}
\tau^{a}R^a_{\mu} - e\gamma^{\mu} \frac{1-\tau_{3}}{2}A_{\mu})\psi_i
= 0\label{sp01}
\end{eqnarray}

which may be also written in the form
\begin{eqnarray}
 \{\gamma^{\mu}(\iota \partial_{\mu} - V_{\mu}) - M - S\}\psi_i & = & 0\nonumber
\end{eqnarray}
with the relativistic fields $S(x)$ and $V_{\mu}(x)$ defined,
respectively by
\begin{eqnarray}
S = g_{\sigma}\sigma\label{sp01}
\end{eqnarray}
\begin{eqnarray}
V_{\mu} = g_{\omega}\omega_{\mu} + g_{\rho}\tau^{a}R^a_{\mu} +
e\frac{1-\tau_{3}}{2}A_{\mu}\label{vp01}
\end{eqnarray}

Similarly one derives the equations of motions for the meson fields.
For the scalar $\sigma$-meson one obtains the simple Klein-Gordon
equation, whereas in the case of vector mesons $\omega$ and $\rho$,
one obtains the Proca equations. However, using the Lorentz guage
for the vector mesons ($\partial_{\upsilon}\omega^{\upsilon} = 0$
etc.) the Proca equations also can be transformed as Klein-Gordon
equations. Thus for the mesons and photons the field equations are
given by
\begin{eqnarray}
(\Box + m_{\sigma}^{2}) \sigma = -g_{\sigma}\rho_s
-g_{2}\sigma^{2}-g_{3}\sigma^{3}
\end{eqnarray}
\begin{eqnarray}
(\Box + m_{\omega}^{2}) \omega^{\mu} = g_{\omega}\rho^{\mu}  -
c_{3}\omega_{\upsilon}\omega^{\upsilon}\omega^{\mu}
\end{eqnarray}
\begin{eqnarray}
(\Box + m_{\rho}^{2}) {R}^{a,\mu} = g_{\rho}\rho_a^{\mu}
\end{eqnarray}
\begin{eqnarray}
\Box A^{\mu}\,\, = e\rho_{Pr}^{\mu}
\end{eqnarray}
where
\begin{eqnarray}
\Box \equiv \frac{\partial ^{2}}{\partial t^{2}} - \nabla^{2},
\,\,\,\,\,\,\,\, (\hbar = c = 1)\nonumber
\end{eqnarray}
As mentioned earlier, the various densities appearing as source
terms in the equations for the meson-fields are obtained in the
mean-field and no-sea approximations whereby the nucleon field
operator $\hat{\psi}$ is expanded in terms of single particle wave
functions $\psi_i$. Thus the scalar density $\rho_{s}$, the nucleon
current density $\rho_{\mu}$, the isovector current density
$\rho_{a,\mu}$ and the electromagnetic current density
$\rho_{Pr,\mu}$ are simply given by
\begin{eqnarray}
\rho_{s}= \sum^{A}_{i=1} w_i\bar\psi_i \psi_i \label{t004}
\end{eqnarray}
\begin{eqnarray}
\rho_{\mu}= \sum^{A}_{i=1} w_i\bar\psi_i\gamma_{\mu} \psi_i
\end{eqnarray}
\begin{eqnarray}
\rho_{a,\mu}= \sum^{A}_{i=1} w_i\bar\psi_i\gamma_{\mu}\tau_a \psi_i
\end{eqnarray}
\begin{eqnarray}
\rho_{Pr,\mu}= \sum^{A}_{i=1}
w_i\bar\psi_i\frac{1-\tau_{3}}{2}\gamma_{\mu}\psi_i\label{t005}
\end{eqnarray}
Here the subscript `Pr' has been used to denote protons. Also  we
have introduced the occupation weights $w_i$ ($0\leq w_i \leq 1$)
which would facilitate the treatment of open shell nuclei. In the
context of pairing correlations, these weights are identical to the
BCS factors $v_{i}^{2}$ (occupation probabilities). The RMF
equations  given by (4), and (7) - (10) are a set of coupled
equations for the nucleon fields, meson fields and the Coulomb field
which are solved by iteration after applying suitable
approximations.

\subsection{Total Energy of the System}
The relation for the energy of the system within the present meson
field theory is derived \cite{serot} from the Hamiltonian density
$\cal H$ using the relation
\begin{eqnarray}
{E_{MF}}&= &{\int d^{3}r}{\cal H}
\end{eqnarray}
where the Hamiltonian density $\cal H$ is defined in terms of $\cal
L$ as
\begin{eqnarray} \cal H & = & \sum_{q_{j}}\pi_{q_{j}} \dot{q}_{j}-
\cal{L}
\end{eqnarray}
Here the canonical momentum $\pi_{q_{j}}$ for the various fields
($q_j$ = $\psi_i,\, \sigma,\, \omega^{\mu}$, $R^{a}_{\mu}$ and
$A^{\mu}$) is given by
\begin{eqnarray}
\pi_{q_{j}} & = & \frac{\partial \cal{L}}{\partial \dot{q}_{j}}
\end{eqnarray}

The mean-field energy $E_{MF}$ obtained above is the major portion
of the energy of a nuclear system. The total energy of the system is
given by
\begin{equation}
E_{total}\,=\,E_{MF}\,+\,E_{Pair}\,-\,E_{CM}\,-\,Z\,M_p\,-N\,M_n
\label{t1009}
\end{equation}
wherein $E_{Pair}$ and $E_{CM}$ are the pairing energy and the
center-of-mass energy respectively.

In order to calculate the pairing energy \cite{ring3,lane} we employ
the state dependent gap equation for the single particle states
wherein for the pairing interaction we have used a delta-function
force. The evaluation of the pairing energy
\begin{eqnarray}
 E_{Pair} = - \sum_{j>0}\Delta_{j}u_{j}v_{j}
\end{eqnarray}
contribution \cite{ring3,lane} involves finally the calculation of
the pairing interaction matrix elements and the single particle
pairing gaps $\Delta_{j}$ along with the occupation probabilities
$v^2_{j}$ etc. For the correction to the center of mass motion we
use the nonrelativistic expression \cite{pgr2}
\begin{eqnarray}
E_{CM} = \frac{\langle P^2_{CM}\rangle} {2M \, A}
\end{eqnarray}
where $P_{CM}$ is the classical center of mass and $A$ is the total
number of nucleons in the nucleus. A simple harmonic oscillator
shell model estimate given by $E_{CM}$ = $\frac{3}{4} 41$ MeV
$A^{-1/3}$ provides a reasonably good approximate description
\cite{pgr2} and has been used in our present study.

\subsection{Specialization to Spherically Symmetric Nuclei}

In the case of spherical nuclei, i.e. the systems which have
rotational symmetry, the potential of the nucleon and the sources of
the meson fields depend only on the radial coordinate r. The field
equations obtained above are further simplified due to spherical
symmetry. In this case, the spinors $\psi_i$ describing nucleons are
characterized as usual in terms of the single particle angular
momentum quantum numbers $j_i$ and $m_i$ and expressed in terms of
radial functions $G_i(r)$ and $F_i(r)$ for the upper and lower
components, respectively, and the spinor
spherical harmonics ${\mathcal Y}_{j_i\, l_i\, m_i}$,\\
\begin{equation}
   \psi_i={1 \over r} \,\, \left({i \,\,\, G_i(r) \,\,\,
    {\mathcal Y}_{j_i\, l_i\, m_i}
    \atop{F_i(r) \, \frac{\sigma \,\cdot \, {r}} {r}\, \,
     {\mathcal Y}_{j_i \,l_i\, m_i}}} \right)\,\,\label{t002}
\end{equation}\\
where the spinor spherical harmonic for a given nucleon $i$ with
quantum numbers $j,\,l\,$ and $m$ is defined by
\begin{equation}
{\mathcal Y}_{j\, l\, m} = \sum_{m_l\,m_s}\,\langle \,\frac{1}{2} \,
m_s\,l \, m_l|l\,m\rangle \, Y_{l\,m}(\theta,\,\phi) \,\chi_{m_s}(s)
\end{equation}
The spinors $\psi_i$ satisfy the normalization condition
\begin{eqnarray}
\int\psi_i^{\dag} \psi_i \,d^{3}x=1\nonumber
\end{eqnarray}
which yields
\begin{eqnarray}
\int dr\, {\{|G_i|^2\,+\,|F_i|^2}\}\,=\,1\label{t003}
\end{eqnarray}

For the calculation of pairing energy we employ the state dependent
gap equation \cite{ring3,lane} for the single particle states
\begin{eqnarray}
     \Delta_{j_1}& =&\,-\frac{1}{2}\frac{1}{\sqrt{2j_1+1}}
     \sum_{j_2}\frac{\left<{({j_1}^2)\,0^+\,|V|\,({j_2}^2)\,0^+}\right>}
      {\sqrt{\big(\varepsilon_{j_2}\,-\,\lambda \big)
       ^2\,+\,{\Delta_{j_2}^2}}}\,\,\sqrt{2j_2+1}\,\,\,
       \Delta_{j_2}\,\,\,\label{t01400}
\end{eqnarray}\\
where $\varepsilon_{j}$ are the single particle energies, and
$\lambda$ is the Fermi energy. The particle number  condition is
expressed in terms of the occupation probabilities for the single
particle states through
\begin{eqnarray}
{\sum_{j}} (2j+1)\, v^2_{j}\,=\,N, \nonumber
\end{eqnarray}
where N is the number of particles in the system, and $v^2_{j}$ are
the occupation probabilities given by
\begin{eqnarray}
v^2_{j}\,=\,\frac{1}{2}\bigg(1 -
\frac{\varepsilon_j\,-\,\lambda}{\sqrt{\big(\varepsilon_j\,-\,\lambda
\big)^2\,+\,{\Delta_{j}^2}}}\bigg)
\end{eqnarray}
In our calculations we use for the pairing interaction the
$\delta$-force, that is $V=-V_{0}\delta(r)$ where $V_{0}$ denotes
the strength. The final result for the paring matrix element is
given by
\begin{eqnarray}
\left<{({j_1}^2)\,0^+\,|V|\,({j_2}^2)\,0^+}\right>&
=&\,\frac{V_0}{8\pi}
       (-1)^{l_1+l_2}\,\,\sqrt{(2j_1+1)(2j_2+1)}\,\,I_R\,\,\,
\end{eqnarray}
where $I_R$ is the radial integral having the form
\begin{eqnarray}
   I_R& =&\,\int\,dr \frac{1}{r^2}\,\left(G^\ast_{j_ 1}\, G_{j_2}\,+\,
     F^\ast_{j_ 1}\, F_{j_2}\right)^2
\end{eqnarray}
For the purpose of our RMF+BCS calculations the value of the pairing
interaction strength $V_{0} = 350$ MeV $fm^{3}$ was determined by
obtaining a best fit to the binding energy of $\rm Ni$ isotopes. We
use the same value of $V_{0} = 350$ MeV $fm^{3}$ throughout for our
present studies of all the other nuclei as well. Moreover, the same
strength has been used for both the protons and neutrons.

\subsection{Specialization to Axially Deformed Nuclei}

The relativistic mean field description has been extended for the
deformed nuclei of axially symmetric shapes by Gambhir, Ring and
their collaborators \cite{gambhir} using an expansion method. The
treatment of pairing has been carried out in Ref. \cite{geng1} using
state dependent BCS method \cite{lane} as has been given by Yadav et
al. \cite{yadav,yadav1} for the spherical case. For axially deformed
nuclei the rotational symmetry is no more valid and the total
angular momentum $j$ is no longer a good quantum number.
Nevertheless, the various densities still are invariant with respect
to a rotation around the symmetry axis. Here we have taken the
symmetry axis to be the z-axis. Following Gambhir {\it et al.}
\cite{gambhir}, it is then convenient to employ the cylindrical
coordinates

\begin{equation}
x = r_\bot \cos\varphi ,\quad y = r_\bot \sin\varphi\quad {\rm
and}\quad z.
\end{equation}

The spinor $\psi_i$ with the index $i$ is now labeled by the quantum
numbers
$\Omega_i, \pi_i$ and $t_i$,
where $\Omega_i$ is the eigenvalue of the symmetry operator
$j_{z_{i}}$ (the projection of $j_{i}$ on the z-axis), $\pi_i$
indicates the parity and $t_i$ has been used for the isospin. In
terms of these quantum numbers, the spinor can now be expressed in
the form:
\smallskip

\begin{equation}
\psi_i ({\bf r},t)\ =\ \begin{pmatrix}f_i({\bf r}) \\ ig_i({\bf
r})\end{pmatrix}
= {1\over\sqrt{2\pi}}\begin{pmatrix} f_i^+(z,r_\bot)& e^{i(\Omega_i-1/2)\varphi}\\
 f_i^-(z,r_\bot) & e^{i(\Omega_i+1/2)\varphi}\\
ig_i^+(z,r_\bot) & e^{i(\Omega_i-1/2)\varphi}\\
ig_i^-(z,r_\bot) & e^{i(\Omega_i+1/2)\varphi}
\end{pmatrix}\chi_{t_i}(t)
\end{equation}


\medskip\noindent
Here the four components $f_i^\pm(r_\bot ,z)$ and $g_i^\pm(r_\bot
,z)$ obey the Dirac equations. For the axially symmetric case the
spinors $f_i^\pm(r_\bot ,z)$ and $g_i^\pm(r_\bot ,z)$ are expanded
in terms of the eigenfunctions of a deformed axially symmetric
oscillator potential as has been described in the Refs.
\cite{gambhir} and \cite{geng1}. The pairing gap $\Delta_{k}$
appearing in the Eqn. (19) satisfies the gap equation

\begin{eqnarray}
 \Delta_{k} = \frac{1}{2}
     \sum_{k'>0}\frac{{\bar{V}_{kk'}}\mid \Delta_{k'}\mid}
      {\sqrt{\big(\varepsilon_{k'}\,-\,\lambda \big)
       ^2\,+\,{\Delta_{k'}^2}}}  \,\,
\end{eqnarray}

Here the symbols $\varepsilon_{k'}$ and $\lambda$ denote the single
particle and Fermi energy, whereas the pairing matrix element
$\bar{V}_{kk'}$ for the symmetrically deformed case using the
zero-range $\delta$-force is given by

\begin{eqnarray}
\bar{V}_{ij}\,\, = \,\, <i\bar{i} \mid V \mid j\bar{j}>\, - \,
<i\bar{i} \mid V \mid \bar{j}j> \\
\hspace{0.8cm}= -V_{0}\int d^{3}r
[\psi_{i}^{\dag}\psi_{\bar{i}}^{\dag}\psi_{j}\psi_{\bar{j}} \, - \,
\psi_{i}^{\dag}\psi_{\bar{i}}^{\dag}\psi_{\bar{j}}\psi_{j} ]
\end{eqnarray}

A detailed description of the relativistic mean-field plus state
dependent BCS approach can be found in Refs.
\cite{yadav,yadav1,yadav2,gambhir,geng1,gambhir1,geng2}.

\section{Results and Discussion}

In sec. 3.1 we have described the results of the calculated binding
energies, one- and two-proton separation energies obtained by
employing RMF+BCS approach including deformation degree of freedom
assuming axially symmetric shapes of nuclei (referred to throughout
as deformed RMF approach) to identify the nuclei which satisfy the
criteria (S$_{p}$ $>$ 0 and S$_{2p}$ $<$ 0) in the region 20 $ \leq$
Z $\leq$ 40.

Nuclei satisfying the above criteria have been predicted as
potential candidates for exhibiting two-proton radioactivity. It is
found that barring a few cases  most of the potential candidates for
the two proton radioactivity are well deformed proton rich nuclei as
has been discussed in sec. 3.2. A detailed examination of the
results for the calculated value of the deformation obtained for the
potential candidates in sec. 3.2 shows that, for example, the
candidate $^{48}$Ni has spherical shape, whereas the potential
nuclei $^{60}$Ge and $^{42}$Cr are slightly deformed. These
spherical or near spherical potential candidates are especially
useful in our theoretical studies since these can be described in
detail within the spherical RMF framework in terms of spherical
single particle wave functions and energies. This in turn enables us
to discuss the nuclear structure aspects of these nuclei in greater
details in terms of relativistic mean field (RMF) potential, wave
function, proton single particle spectrum and density distributions
etc. as has been elucidated in sec. 3.2.

In order to obtain the one- and two-proton separation energies using
Eqns. (\ref{eqq1}) and (\ref{eqq2}) given below, systematic
calculations are carried out within the framework of deformed RMF
approach for the even-Z nuclei in the region 20 $ \leq$ Z $\leq$ 40.
Thus for a nucleus with proton and neutron number (Z, N), the
calculations for S$_{p}$ and S$_{2p}$ require the deformed RMF
results for the chain of isotopes with proton number Z, Z-1 and Z-2,
as is evident from the form of expressions (\ref{eqq1}) and
(\ref{eqq2}). In order to proceed towards the proton drip-line and
obtain the S$_{p}$ and S$_{2p}$ values for the proton rich nucleus,
similar calculations are repeated in a systematic way for decreasing
value of N to obtain the results for the nuclei of the isotopic
chain (Z, N-1), (Z, N-2), (Z, N-3)... etc. The calculated deformed
RMF results for the binding energies then yield the desired S$_{p}$
and S$_{2p}$ values up to the proton drip-line and beyond. These
results are then plotted as a function of decreasing mass number A
for different values of Z and analysed to predict the potential
candidates for two-proton radioactivity as has been described in the
subsection below. Calculation for neutron rich nuclei (N $>$ Z) have
not been shown and discussed as they are not relevant to our study
of the phenomenon of two-proton radioactivity.

\subsection{Analysis of separation energy for
the identification of two-proton emitters}

The results of our extensive calculations obtained for the nuclei
constituting the Z = 20 to Z = 40 isotopic chains by employing the
deformed RMF approach have been shown in Fig. 1 as a function of
increasing mass number A for all the isotopic chains of nuclei with
even Z values in the region 20 $ \leq$ Z $\leq$ 40 along with the
available experimental data \cite{audi}.

From the figure it is observed that the calculated binding energy
values are very close to the available experimental data
\cite{audi}. However, due to large scales involved in this figure,
small differences between the calculated and experimental values of
the binding energy (0.1 MeV to 3 MeV) are not distinctly visible in
Fig. 1. The maximum difference of 3 MeV occurs in the case of few
proton rich nuclei located in the vicinity of Z = 26 (Fe) isotopic
chain. This difference is about 1 percent of the total binding
energy of the respective nucleus in that region.

\begin{figure}[th]
\centerline{\psfig{file=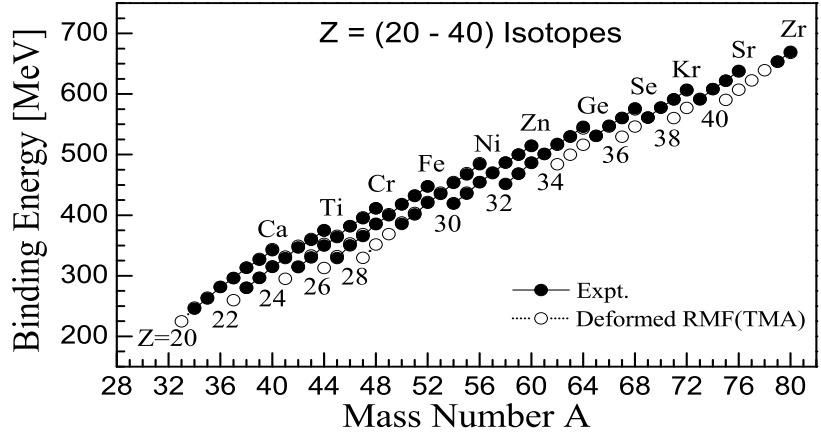,width=11 cm, height=7cm}}
\caption{The results for the binding energy obtained within the
framework of deformed RMF approach using the TMA force parameters
for the nuclei belonging to the isotopic chains of Z = 20 to Z = 40
have been shown as open circles. These calculated results are
compared with the available experimental data shown as filled
circles $^{47}$.}
\end{figure}

Due to close agreement between the theoretical and experimental
results for the binding energy, we can accurately determine one- and
two-proton separation energy values which are very crucial for
making reliable prediction of the two-proton emitters. Thus, the
deformed RMF approach is expected to describe the phenomena of
two-proton radioactivity in the region 20 $\leq$ Z $\leq$ 40 in a
more realistic manner as compared to other theoretical approaches
wherein the accuracies are rather not really large.

As mentioned earlier, since the nuclei satisfying the condition
S$_{p}$ $>$ 0 and S$_{2p}$ $<$ 0 might be the possible candidates
for simultaneous two-proton emission, we obtain reliable and
accurate calculations for the S$_{p}$ and S$_{2p}$ values. In fact
these are obtained from the calculated binding energies using the
expressions,

\begin{eqnarray}
S_{2p}(Z,N) = B(Z,N) - B(Z-2,N) \label{eqq1}
\end{eqnarray}
\begin{eqnarray}
S_{p}(Z,N) = B(Z,N) - B(Z-1,N) \label{eqq2}
\end{eqnarray}

\noindent where B(Z,N) is the binding energy of the nucleus with Z
protons and N neutrons. The calculated results of the one- and
two-proton separation energy for the nuclei belonging to the
isotopic chains of nuclei with Z = 20 to 40 have been displayed in
Figs. 2 and 3, respectively, as a function of decreasing mass number
A. In these figures, the filled circles denote the experimental
value of one- and two-proton separation energies, whereas the open
circles represent the calculated results.

\begin{figure}[th]
\centerline{\psfig{file=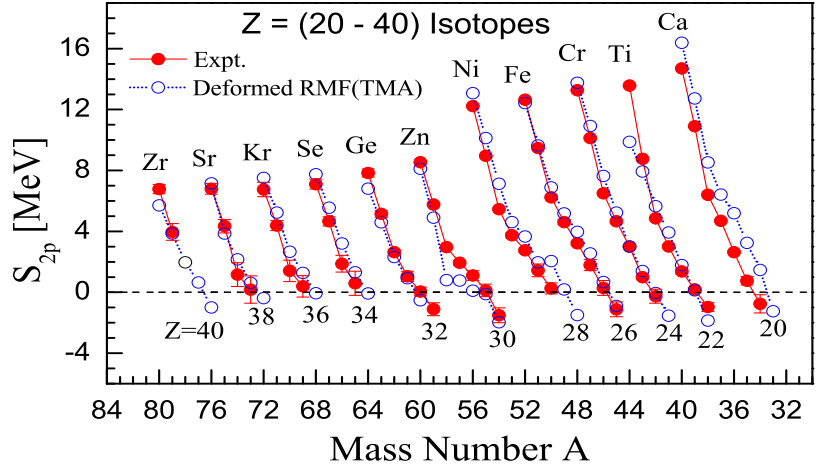,width=11 cm, height=7cm}}
\caption{Calculated results for the two-proton separation energy
S$_{2p}$ obtained in the deformed RMF approach using the TMA force
parameters are shown by open circles. The lines connecting different
isotopes with proton number lying between Z = 20 to Z = 40 have been
drawn to guide the eyes. The filled circles denote the corresponding
available experimental data with error bars for the two-proton
separation energy $^{47}$.}
\end{figure}

A comparison of the results obtained in the deformed RMF approach
with the available experimental data for the one- and two-proton
separation energies shows that in general theoretical results are in
fairly good agreement with the measurements, though for some cases
the calculated values are rather appreciably different from the
experimental data. In a few cases this marked difference between the
calculated results and the data can be attributed to the large
uncertainty in the measurements shown by the error bars in Figs. 2
and 3.

It is seen from Fig. 2 for the two-proton separation energy that
nuclei $^{45}$Fe, $^{48}$Ni and $^{54}$Zn which have been shown to
be two-proton emitters in recent experiments
\cite{Pftzner,Giovinazzo,Blank,Ascher,Dossat} are located beyond the
two-proton drip-line with negative two-proton separation energy
values -0.90 MeV, -1.49 MeV and -1.97 MeV, respectively. Out of
these, nuclei $^{45}$Fe and $^{48}$Ni have positive one-proton
separation energy (S$_{p}$ $>$ 0) and therefore fulfill the criteria
of being a two-proton emitter (S$_{p}$ $>$ 0 and S$_{2p}$ $<$ 0). In
contrast to the above mentioned two nuclei, $^{45}$Fe and $^{48}$Ni,
for the nucleus $^{54}$Zn both one- and two-proton separation
energies are found to be negative. Interestingly, the nucleus
$^{55}$Zn which belongs to the Z = 30 isotopic chain is found to be
a two-proton emitter according to the results of our deformed RMF
calculations.

\begin{figure}[th]
\centerline{\psfig{file=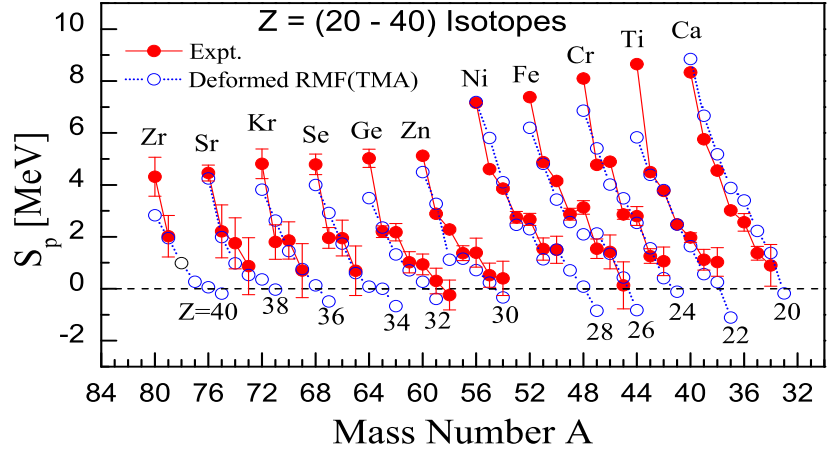,width=11 cm, height=7cm}}
\caption{Calculated results for the one-proton separation energy
S$_{p}$ obtained in our deformed RMF approach using the TMA force
parameters have been depicted by open circles. The lines connecting
the isotopes for the nuclei with proton number ranging from Z = 20
to Z = 40 have been drawn to guide the eyes. Available experimental
data $^{47}$ have been depicted by the filled circles for the
purpose of comparison.}
\end{figure}

Moreover, the result of our extensive calculations  further show
that the nuclei $^{38}$Ti, $^{42}$Cr, $^{60}$Ge, $^{63,64}$Se,
$^{68}$Kr, $^{72}$Sr and $^{76}$Zr satisfy the criteria S$_{p}$ $>$
0 and S$_{2p}$ $<$ 0. These nuclei are, therefore, expected to be
the potential candidates for exhibiting the two-proton
radioactivity. It may be emphasized that the experimental data
\cite{audi} for the separation energies for the nuclei identified
above (with the exception of $^{55}$Zn and $^{60}$Ge) are consistent
with the criteria S$_{p}$ $>$ 0 and S$_{2p}$ $<$ 0. The difference
between the calculated and measured results for nuclei $^{55}$Zn and
$^{60}$Ge can be attributed to the appreciable errors in the
experimental binding energies \cite{audi}. Results of the one- and
two-proton separation energies for the nuclei identified above along
with the available experimental data have been listed in Table 3 to
facilitate an easy comparison.

\begin{table}[h]

\tbl{One- and two-proton separation energies, S$_{p}$ and S$_{2p}$,
respectively, and corresponding available experimental data for the
predicted two-proton emitters.}
{\begin{tabular}{|c|c|c|c|c|c|c|c|c|c|c|}
 \hline
 \multicolumn{1}{|c|}{}&
 \multicolumn{2}{c|}{Theory}&
 \multicolumn{4}{c|}{Expt.}\\
\cline{2-7} \multicolumn{1}{|c}{Nucleus}&
\multicolumn{1}{|c|}{S$_{p}$}&\multicolumn{1}{|c|}{S$_{2p}$}&
\multicolumn{1}{c|}{S$_{p}$}& \multicolumn{1}{|c|}{Error
(S$_{p}$)}&\multicolumn{1}{c|}{S$_{2p}$}&
\multicolumn{1}{c|}{Error (S$_{2p}$)}\\
\multicolumn{1}{|c}{}&
\multicolumn{1}{|c|}{(MeV)}&\multicolumn{1}{|c|}{(MeV)}&
\multicolumn{1}{c|}{(MeV)}&
\multicolumn{1}{|c|}{(MeV)}&\multicolumn{1}{c|}{(MeV)}&
\multicolumn{1}{c|}{(MeV)}\\
\hline
$^{38}$Ti&0.25&-1.86&1.03&0.55&-0.96&0.29\\
$^{42}$Cr&0.41&-0.05&1.06&0.55&-0.26&0.46\\
$^{45}$Fe&0.43&-0.90&0.13&0.90&-1.12&0.49\\
$^{48}$Ni&0.08&-1.49&-&-&-&-\\
$^{55}$Zn&0.24&-0.12&0.52&0.47&0.12&0.41\\
$^{60}$Ge&0.27&-0.52&0.94&0.40&0.05&0.28\\
$^{63}$Se&0.00&-0.96&-&-&-&-\\
$^{64}$Se&0.08&-0.08&-&-&-&-\\
$^{68}$Kr&0.13&-0.06&-&-&-&-\\
$^{72}$Sr&0.36&-0.38&-&-&-&-\\
$^{76}$Zr&0.06&-1.00&-&-&-&-\\
\hline
\end{tabular}}
\end{table}
\vspace{0.1cm}

\noindent It may be mentioned that the separation energy values near
the drip-line are close to zero and therefore the sign of the
separation energy value is very sensitive and at times model
dependent.

In order to check the force parameter dependence, we have carried out the
RMF+BCS calculations for some nuclei using the NL-SH force
parameters\cite{sharma} which are equally popular for the
relativistic mean-field calculations. It is found that generally our
RMF+BCS results for the two force parameters are similar though
there are always some differences in finer details. However, it is
worthwhile to note that our predictions of potential two-proton
emitters are not affected if we use NL-SH force
parameters\cite{sharma} instead of TMA force
parameters\cite{suga,suga2} as evident from Table 4.

\begin{table} [h]
\tbl{Comparison between one-proton separation energy (S$_{p}$) and
two-proton separation energy (S$_{2p}$) obtained by using TMA and
NL-SH force parameters in respect of some selected nuclei.}
{\begin{tabular}{|c|c|c|c|c|c|c|c|c|c|c|}
 \hline
 \multicolumn{1}{|c|}{}&
 \multicolumn{3}{c|}{RMF+BCS(TMA)}&
 \multicolumn{3}{c|}{RMF+BCS(NL-SH)}\\
\cline{2-7} \multicolumn{1}{|c}{Nucleus}&
\multicolumn{1}{|c|}{S$_{p}$}&\multicolumn{1}{|c|}{S$_{2p}$}&\multicolumn{1}{|c|}{Whether
2p emission}& \multicolumn{1}{c|}{S$_{p}$}&
\multicolumn{1}{c|}{S$_{2p}$}&\multicolumn{1}{|c|}{Whether 2p emission}\\
\multicolumn{1}{|c}{}&
\multicolumn{1}{|c|}{(MeV)}&\multicolumn{1}{|c|}{(MeV)}&
\multicolumn{1}{c|}{condition
fulfilled}&\multicolumn{1}{c|}{(MeV)}&\multicolumn{1}{c|}{(MeV)}&
\multicolumn{1}{c|}{condition fulfilled}\\
\hline
$^{45}$Fe&0.43&-0.90&Yes&0.88&-0.51&Yes\\
$^{48}$Ni&0.08&-1.49&Yes&0.54&-1.19&Yes\\
$^{54}$Zn&-0.34&-1.97&No&-1.79&-4.31&No\\
\hline
\end{tabular}}
\end{table}

\subsection{Representative Examples of Two-Proton Emitters Located
Beyond the Two-Proton Drip-Line}

The results of the calculated quadrupole deformation values of the
predicted two-proton emitters and their respective daughter nuclei
are summarized below Table 5, especially to highlight and identify
the nuclei which are spherical in shape as these may be studied in
greater detail within the framework of spherical RMF.

\begin{table} [h]
\tbl{Results of the quadrupole deformation parameter for the matter
density distribution $\beta_{2m}$ for the predicted two-proton
emitters and their daughter nuclei.} {\begin{tabular}{|c|c|c|c|}
 \hline  \multicolumn{1}{|c|}{Parent Nucleus}&
 \multicolumn{1}{|c|}{$\beta_{2m}$}&
 \multicolumn{1}{|c|}{Daughter Nucleus}&
 \multicolumn{1}{|c|}{$\beta_{2m}$}\\
 \hline
$^{38}$Ti&0.22&$^{36}$Ca&0.00\\
$^{42}$Cr&-0.17&$^{40}$Ti&-0.15\\
$^{45}$Fe&0.00&$^{43}$Cr&-0.07\\
$^{48}$Ni&0.00&$^{46}$Fe&0.00\\
$^{55}$Zn&0.25&$^{53}$Ni&0.14\\
$^{60}$Ge&0.13&$^{58}$Zn&0.10\\
$^{63}$Se&-0.23&$^{61}$Ge&0.19\\
$^{64}$Se&0.24&$^{62}$Ge&0.23\\
$^{68}$Kr&-0.29&$^{66}$Se&-0.25\\
$^{72}$Sr&-0.27&$^{70}$Kr&-0.31\\
$^{76}$Zr&-0.33&$^{74}$Sr&-0.35\\
 \hline
\end{tabular}}
\end{table}

\vspace{0.1cm}

\noindent From the table it is seen that nuclei $^{45}$Fe, $^{48}$Ni
and $^{60}$Ge as well as their corresponding daughter nuclei have
spherical or near spherical shapes. In contrast, the nuclei
$^{55}$Zn and $^{64}$Se, and also their respective daughter nuclei
are seen to have prolate deformation, whereas the nuclei $^{42}$Cr,
$^{68}$Kr, $^{72}$Sr and $^{76}$Zr, and their respective daughter
nuclei have oblate shapes. From these results, it is evidently seen
that besides the nuclei $^{38}$Ti and $^{63}$Se, all other
identified two-proton emitters and their corresponding daughter
nuclei have similar shapes. Thus, it appears that the shape of the
two-proton emitter nucleus is almost preserved in the process of
two-proton emissions.

From amongst the predicted two-proton radioactive nuclei mentioned
above, we have chosen the nuclei $^{42}$Cr, $^{48}$Ni and $^{60}$Ge
which have spherical or near spherical shapes as can be seen from
Table 4, to describe their detailed properties within the framework
of the spherical RMF approach in order to obtain greater insight
into the structural properties of these two-proton emitters.

It should be emphasized that though the nuclei $^{42}$Cr and
$^{60}$Ge are slightly deformed, these have been treated here within
the spherical RMF approach only as an approximation. On the other
hand not only $^{48}$Ni is spherical but most of the Ni isotopes are
found to be spherical in shape indicating Z = 28 remains a good
magic number for all the isotopes with neutron number ranging from N
= 20 to N = 70. It should also be stated that the nuclei $^{42}$Cr,
$^{48}$Ni and $^{60}$Ge which have been identified as two-proton
emitters are extremely proton rich nuclei located beyond the
two-proton drip-line. Since the Ni isotopes are found to be
spherical, these are ideally suited to be studied within the
spherical RMF approach without any approximation. Thus these
isotopes have been studied in rather great detail in order to show
that the physical characteristics exhibited by bound proton rich
isotopes remain intact even beyond the two-proton drip-line. It is
found that the nucleus $^{48}$Ni which lies beyond the two-proton
drip-line and thus is unbound, due to the centrifugal and Coulomb
barrier develops finite life time and eventually decays via
two-proton radioactivity. In the following we have also presented
some detailed results for the proton rich $^{48-56}$Ni isotopes to
demonstrate a systematic variation in the mean-field potential, wave
function, energy of the single particle states and the proton
density with increasing neutron number as we move towards the line
of stability. The most interesting result of these study of Ni
isotopes is that the highest proton single particle state
1f$_{7/2}$, for example, which remains completely filled and bound
for all the Ni isotopes up to the proton drip-line, preserves its
characteristics even beyond the drip-line when the nucleus $^{48}$Ni
becomes unbound as it exhibits negative two-proton separation
energy. This explains the long life time of this isotope even when
it lies beyond the two-proton drip-line and eventually decays. This
is manifested in the form of two proton radioactivity. This
conclusion is reinforced by observing a similar characteristics of
the radial density distribution for the $^{48}$Ni isotope as
compared to those of the bound isotopes $^{50-56}$Ni. The shell
closure or magicity of the $^{48}$Ni isotope also remains preserved
as for the bound isotopes $^{50-56}$Ni. These are significant
characteristics explaining why the nucleus $^{48}$Ni even while
lying beyond the two-proton drip-line has positive one-proton
separation energy and decays via two-proton radioactive mode.

\subsubsection{The nucleus $^{42}_{24}$Cr$_{18}$}

\begin{figure}[th]
\centerline{\psfig{file=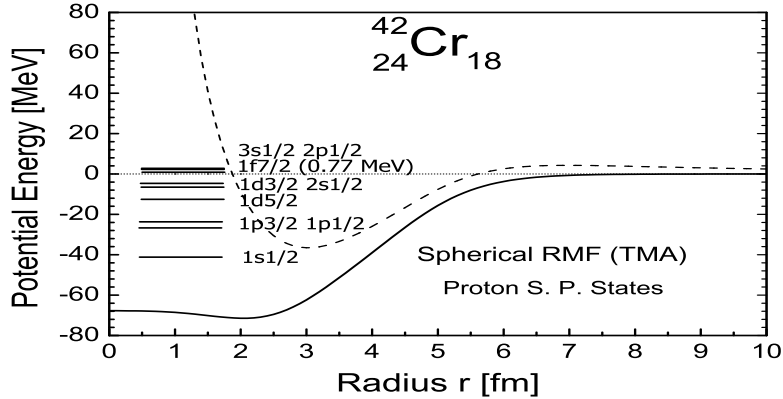,width=11cm,height=6cm}}
\caption{The RMF potential energy (sum of the scalar and vector
potentials) for the nucleus $^{42}_{24}$Cr$_{18}$ as a function of
radius is shown by the solid line. The long dashed line represents
the sum of RMF potential energy and the centrifugal barrier energy
for the proton resonant state 1f$_{7/2}$.}
\end{figure}

The predicted two-proton emitter nucleus $^{42}_{24}$Cr$_{18}$ is
found to be somewhat deformed with $\beta_{2m}$ = -0.17. However, as
mentioned above, in order to learn its detailed structure in terms
of spherical single particle energies and wave functions, we have
studied it within the framework of the spherical RMF approach. This
approximate treatment, though not fully justified, is expected to
shed light as regard to its finite life time even though it is
located beyond the two-proton drip-line.

As mentioned earlier, this nucleus acquires long life time against
two-proton decay due to the combined barrier provided by the Coulomb
and centrifugal effects, even though it has negative two proton
separation energy. In order to demonstrate the physical situation of
proton rich $^{42}_{24}$Cr$_{18}$ nucleus, we have plotted in Fig. 4
the RMF potential and the single particle energy spectrum for the
bound proton single particle states. The figure also depicts a few
positive energy proton states in the continuum including the
resonant states 1f$_{7/2}$ at energy 0.77 MeV. We have also shown in
Fig. 4 the total mean-field potential for the resonant $1f_{7/2}$
state, obtained by adding the centrifugal potential energy.

\begin{figure}[th]
\centerline{\psfig{file=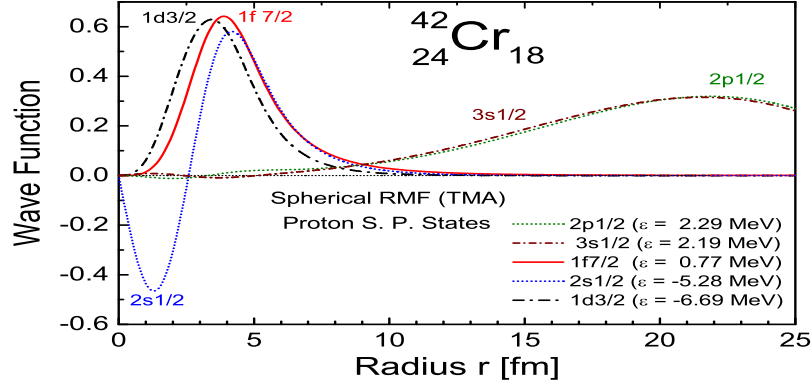,width=11cm,height=6cm}}
\caption{Radial wave functions of a few representative proton single
particle states with energy close to the Fermi surface for the
nucleus $^{42}_{24}$Cr$_{18}$. The resonant state (1f$_{7/2}$)
similar to the bound states (2s$_{1/2}$ and 1$d_{3/2}$) is mostly
confined within the potential region.}
\end{figure}

It may be emphasized that besides the resonant state 1f$_{7/2}$,
other positive energy proton single particle states do not play any
significant role in contributing to the total pairing energy as only
this state has substantial overlap with the bound states near the
Fermi level. This can be inferred from  Fig. 5 wherein we have
displayed the radial wave functions of some of the proton single
particle states close to the Fermi surface, the proton Fermi energy
being $\lambda_p\, =\,0.80$ MeV. These include the bound  states
2s$_{1/2}$ and 1$d_{3/2}$,  and the continuum states 2p$_{1/2}$ and
3$s_{1/2}$, in addition to the resonant state 1f$_{7/2}$.

The wave function for the 1f$_{7/2}$ state plotted in Fig. 5 is
clearly seen to be confined within a radial range of about 7 fm and
has a decaying component outside this region, characterizing a
resonant state. In contrast, the main part of the wave function for
the non-resonant states, e.g. 2p$_{1/2}$  and 3s$_{1/2}$, is seen to
be spread over outside the potential region, though a small part is
also contained inside the potential range. This type of states thus
has a poorer overlap with the bound states near the Fermi surface
leading to small value for the pairing gap $\Delta_{j}$. Further,
the positive energy states lying much higher from the Fermi level,
for example, 1g$_{7/2}$, 1g$_{9/2}$ etc.  have a negligible
contribution to the total pairing energy of the system.

\begin{figure}[th]
\centerline{\psfig{file=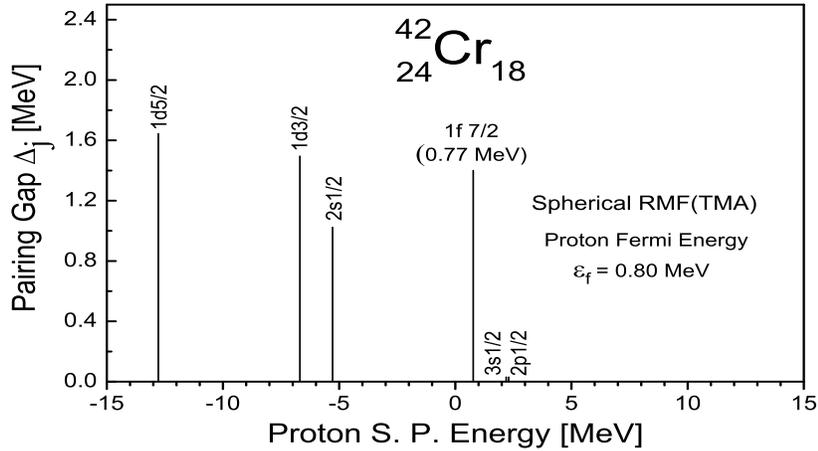,width=11cm,height=7cm}}
\caption{Pairing gap energy $\Delta_{j}$ of proton single particle
states with energy close to the Fermi surface for the nucleus
$^{42}_{32}$Cr$_{18}$. The proton resonant single particle state
1f$_{7/2}$ at energy 0.77 MeV has the gap energy of about 1 MeV
which is close to that of bound states $1d_{5/2}$, $1d_{3/2}$ and
$2s_{1/2}$. In this respect the resonant state $1f_{7/2}$ behaves
similar to that of a bound state such as $1d_{5/2}$, $1d_{3/2}$,
$2s_{1/2}$ etc.}
\end{figure}

This is also clearly evident from Fig. 6 which shows the calculated
single particle pairing gap energy $\Delta_j$ for some of the proton
single particle states in the nucleus $^{42}_{24}$Cr$_{18}$.
However, we have not shown in the figure the single particle states
having negligibly small $\Delta_j$ values as these do not contribute
significantly to the total pairing energy. One observes indeed in
Fig. 6 that the proton resonant single particle state 1f$_{7/2}$ at
energy 0.77 MeV has the pairing gap energy of about 1 MeV which is
close to that of the bound states $1d_{5/2}$, $1d_{3/2}$ and
$2s_{1/2}$. Also, Fig. 6 shows that the pairing gap value for the
non-resonant states $2p_{1/2}$ and $3s_{1/2}$ lying in the continuum
is negligibly small.

From the characteristics of the proton resonant 1f$_{7/2}$ single
particle state, it is evidently clear that this state behaves
similar to a bound state. This property also enables the nucleus
$^{42}_{24}$Cr$_{18}$ to have finite life time even with negative
two-proton separation energy while lying beyond the two-proton
drip-line as is elucidated below. In the nucleus
$^{42}_{24}$Cr$_{18}$, the last 4 protons outside the closed s-d
shells occupy the 1f$_{7/2}$ single particle state. If we consider
the neighboring isotones of $^{42}_{24}$Cr$_{18}$, that is
$^{38}_{20}$Ca$_{18}$ and $^{40}_{22}$Ti$_{18}$ with increasing
number of protons in the 1f$_{7/2}$ proton single particle state,
calculations show that while the nucleus $^{38}_{20}$Ca$_{18}$ for
which the 1f$_{7/2}$ state is completely empty, is bound and has
spherical shape, the nucleus $^{40}_{22}$Ti$_{18}$ is also bound
though slightly deformed. On further addition of two more protons to
$^{40}_{22}$Ti$_{18}$, the next isotone $^{42}_{24}$Cr$_{18}$ is
formed in which now the 1f$_{7/2}$ resonant state is occupied by
four protons. Due to the resonant nature of the proton single
particle  1f$_{7/2}$ state the pairing energy is increased and the
nucleus tends to remain bound, whereas the Coulomb interactions
amongst the increasing number of protons acts in a disruptive way to
make the nucleus unbound with negative two-proton separation energy.
However, the centrifugal barrier provided by high angular momentum
resonant  1f$_{7/2}$ state together with the Coulomb barrier enables
the nucleus $^{42}_{24}$Cr$_{18}$ to gain finite life time, and
eventually it decays by two proton emission. It should be emphasized
that the contribution of pairing energy plays an important role for
the stability of nuclei and consequently in deciding the position of
the neutron and proton drip-lines. Also it is remarkable that in
contrast to the proton rich nucleus such as $^{42}_{24}$Cr$_{18}$,
if we consider the neutron rich nucleus like $^{60}_{20}$Ca$_{40}$,
it is found that further addition of neutrons while approaching the
extremely neutron rich nucleus $^{70}_{20}$Ca$_{50}$, the single
particle neutron states 3s$_{1/2}$, 1g$_{9/2}$, 2d$_{5/2}$ and
2d$_{3/2}$ which lie near the Fermi level gradually come down close
to zero energy, and subsequently the 1g$_{9/2}$ and 3s$_{1/2}$
neutron single particle states even become bound states. This helps
in accommodating more and more neutrons which are just bound. In the
case of neutrons we do not have the disruptive Coulomb force
anymore. Similar explanation holds good for other two-proton
emitters like $^{38}_{22}$Ti$_{16}$ and $^{60}_{32}$Ge$_{28}$.

\subsubsection{The nucleus $^{48}_{28}$Ni$_{20}$}

\begin{figure}[th]
\centerline{\psfig{file=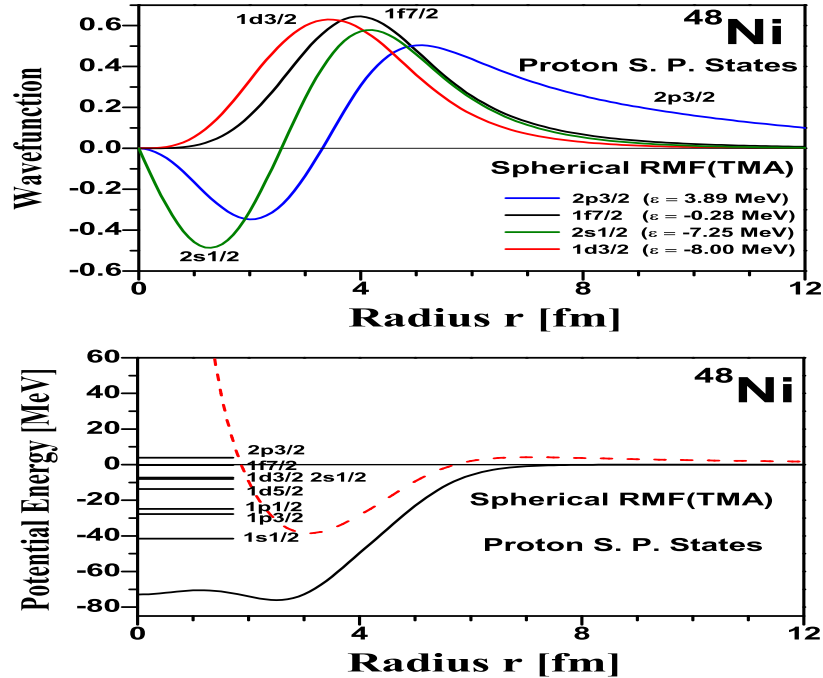,width=11cm,height=9cm}} \caption{Upper
panel: Radial wave functions of a few representative  proton single
particle states for the nucleus $^{48}\rm{Ni}$. It is seen that the
bound single particle states such as $1f_{7/2}$, $2s_{1/2}$ and
$1d_{3/2}$ are mostly confined inside the potential region. In
contrast the $2p_{3/2}$ state which is not a bound state has a large
spread outside the potential region as well.  Lower panel: The
potential energy (sum of the scalar and vector potentials), for the
nucleus $^{48}${Ni} as a function of radius is shown by the solid
line. The dashed line represents the sum of potential energy and the
centrifugal barrier energy for the proton single particle state
$1f_{7/2}$. The effective potential indeed prohibits a rapid decay
of $^{48}${Ni} even though this nucleus is unbound and lies beyond
the two-proton drip-line.}
\end{figure}

In order to illustrate the case of proton rich $^{48}$Ni nucleus, we
have plotted in Fig. 7, the RMF potential and the wave functions of
its few representative proton single particle states in the lower
and upper panels respectively. The lower panel also shows the
spectrum of bound proton single particle states along with the
positive energy state 2p$_{3/2}$ in the continuum. We have also
depicted in lower panel of Fig. 7 the total mean-field potential for
the highest bound and fully occupied proton single particle state
1f$_{7/2}$, obtained by adding the centrifugal potential energy. The
combined effect of Coulomb barrier and centrifugal barrier prevents
the protons from quickly leaving the proton rich nucleus $^{48}$Ni
located beyond the two-proton drip-line. The delay associated with
the tunneling process allows for the observation of two-proton
radioactivity. In the upper panel of Fig. 7 we have displayed the
radial wave functions of some proton single particle states, the
proton Fermi energy being $\lambda_p$ = 2.86 MeV. These include the
bound 1d$_{3/2}$, 2s$_{1/2}$, $1f_{7/2}$ and the continuum
2p$_{3/2}$ proton single particle state.

It is seen that the radial wave function of the proton single
particle state 1f$_{7/2}$ remains mainly confined to the region of
the potential well. In contrast, the wave function for a typical
continuum state 2p$_{3/2}$ is spread over to large distances outside
of the potential region. Therefore, the important outcome from the
above discussions is that the highest proton single particle state
1f$_{7/2}$ remains completely occupied and bound even for the
nucleus $^{48}$Ni lying beyond the two-proton drip-line with
negative two-proton separation energy.

Since $^{48-56}${Ni} isotopes have the spherical shape, it would be
interesting to employ spherical RMF approach to investigate the
behaviour of single particle spectrum, potential, wave function and
density distribution of Ni isotopes with increasing Z/N ratio. With
this in view, first we plot in Fig. 8 the RMF potential (lower
panel), and radial wave function of the proton single particle state
1f$_{7/2}$ for the $^{48-56}${Ni} isotopes as a function of radius
in two different scales (middle and upper panel). We note from the
lower panel of Fig. 8 that although the two-proton separation energy
of nucleus $^{48}$Ni is negative, its potential behaves similar to
the $^{50-56}$Ni isotopes which have positive two-proton separation
energy.

\begin{figure}[th]
\centerline{\psfig{file=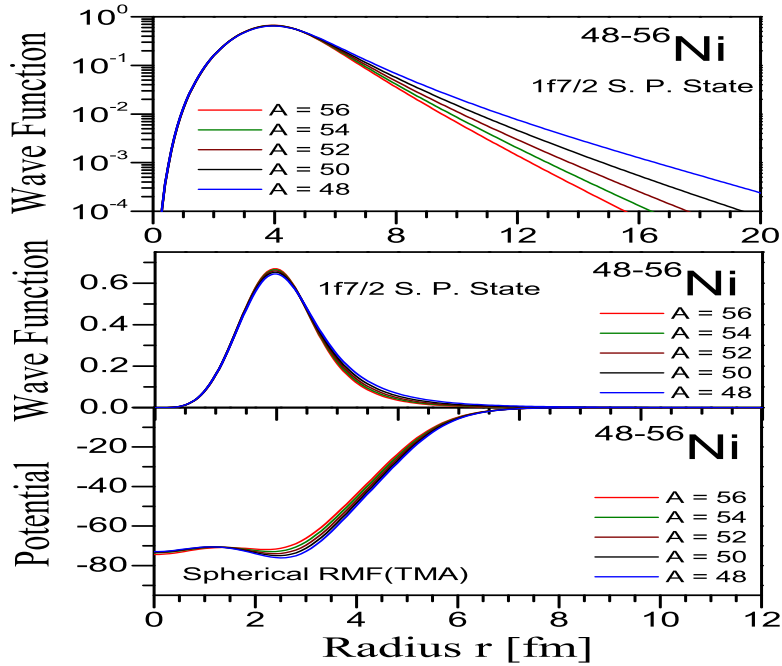,width=11cm,height=9cm}}
\caption{Lower Panel: The RMF potential energy (sum of the scalar
and vector potentials), for the isotopes $^{48-56}$Ni as a function
of radial distance is shown by the solid line. Middle Panel: Radial
wave functions (in a linear scale) of proton single particle state
1f$_{7/2}$ for the isotopes $^{48-56}$Ni. Upper Panel: Radial wave
functions of proton single particle state 1f$_{7/2}$ for the
isotopes $^{48-56}$Ni shown in a logarithmic scale in order to
demonstrate the difference in the wave function for various isotopes
at large distances. It is remarkable that within the potential
region this difference is indeed not large.}
\end{figure}

Moreover, it is clearly seen that similar to the case of
$^{50-56}$Ni isotopes, the proton single particle state 1f$_{7/2}$
in the nucleus $^{48}$Ni remains confined to its potential region.
In order to demonstrate the difference in the wave function for
various isotopes at large distances, we have plotted in the upper
panel of Fig. 8 the radial wave function of proton single particle
state 1f$_{7/2}$ for the $^{48-56}$Ni isotopes in logarithmic scale.
It is seen that the wave function of the proton single particle
state 1f$_{7/2}$ starts spreading slightly outside the potential
region with increasing Z/N ratio. Nevertheless, a large part of the
wave functions is contained inside the potential range. From these
results it is clear that the nucleus $^{48}$Ni lying beyond the
two-proton drip-line behaves similar to the $^{50-56}$Ni isotopes
for which the two-proton separation energy is positive.

\begin{figure}[th]
\centerline{\psfig{file=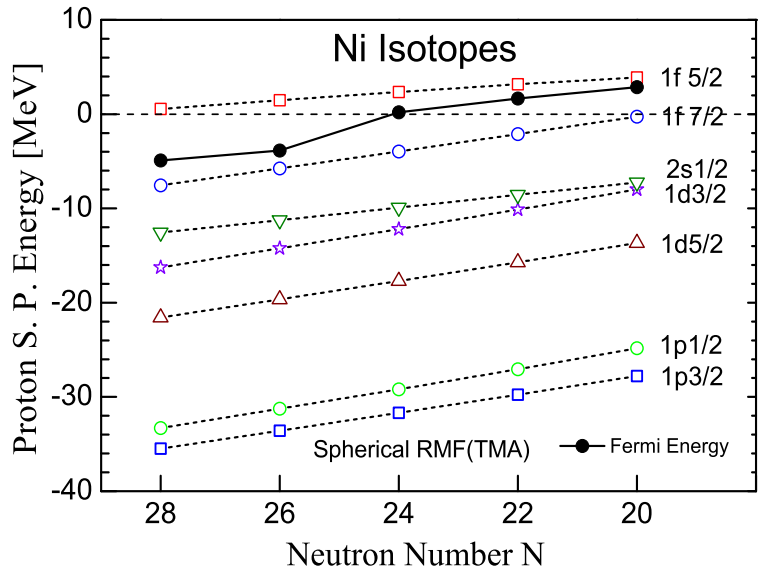,width=11cm,height=8cm}}
\caption{Variation of the proton single particle energies obtained
in the spherical RMF+BCS calculations with the TMA force for the
$^{48-56}$Ni isotopes with decreasing neutron number N (or in other
words with increasing Z/N ratio). Position of the proton Fermi level
has been shown by solid circles. The energy levels have been
connected by dashed line only to guide the eyes. It is interesting
to observe that the nucleus $^{48}$Ni being unbound preserves the
large gap between the 1$f_{5/2}$ and 1$f_{7/2}$ proton single
particle states and the magicity appears to remain intact. This
characteristics is unusual for an unbound nucleus.}
\end{figure}

In Fig. 9 we have shown the variation of the proton single particle
levels and position of the Fermi level of $^{48-56}$Ni isotopes with
decreasing neutron number N (or in other words with increasing Z/N
ratio). It is readily seen from the figure that the existence of
large energy gap between proton single particle levels 1$f_{7/2}$
and 1$f_{5/2}$ explains the tradition proton shell closure at Z = 28
in $^{48-56}$Ni isotopes. The proton Fermi energy which lies at
$\epsilon_f$ = -4.91 MeV in the $^{56}$Ni nucleus moves to
$\epsilon_f$ = 2.86 MeV in the nucleus $^{48}$Ni having maximum Z/N
value.

It is evident from the Fig. 9 that despite the fact that the nucleus
$^{48}$Ni is located beyond the two-proton drip-line, the energy gap
between proton single particle levels 1$f_{7/2}$ and 1$f_{5/2}$
remains significant enough to maintain the Z = 28 shell closure.
This is due to the fact that the highest proton single particle
state 1f$_{7/2}$ preserves its characteristics beyond the two-proton
drip-line. A similar conclusion can be drawn from the variation of
the proton density distribution as a function of radial distances.
Result of such a variation have been depicted in Fig. 10 for Ni
isotopes with N = 20 to 28. As the proton shell in $^{48-56}$Ni
isotopes remains closed we observe sharply falling asymptotic
density distribution for these isotopes. It is remarkable that the
proton density distribution of the unbound $^{48}$Ni isotope has the
radial dependence similar to the other isotopes $^{50-56}$Ni which
are bound. From the various results presented for the Ni isotopes,
it is observed that the unbound proton rich nucleus preserves
properties very similar to the other bound states and this enables
it to have finite life time and decay by emission of two protons.

\begin{figure}[th]
\centerline{\psfig{file=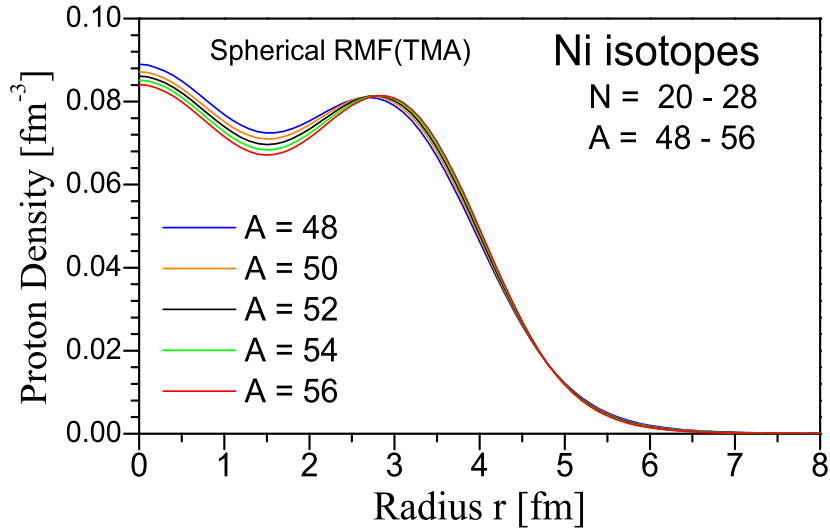,width=11cm,height=8cm}}
\caption{The radial density dependence of the proton density
distributions for the $^{48-56}$Ni isotopes obtained in spherical
RMF approach. The interesting result is that the proton density
distribution of the unbound $^{48}$Ni isotope has the radial
dependence similar to the other isotopes $^{50-56}$Ni which are
bound, this causes the unbound nucleus $^{48}$Ni to have finite
decay time.}
\end{figure}

\subsubsection{The nucleus $^{60}_{32}$Ge$_{28}$}

\begin{figure}[th]
\centerline{\psfig{file=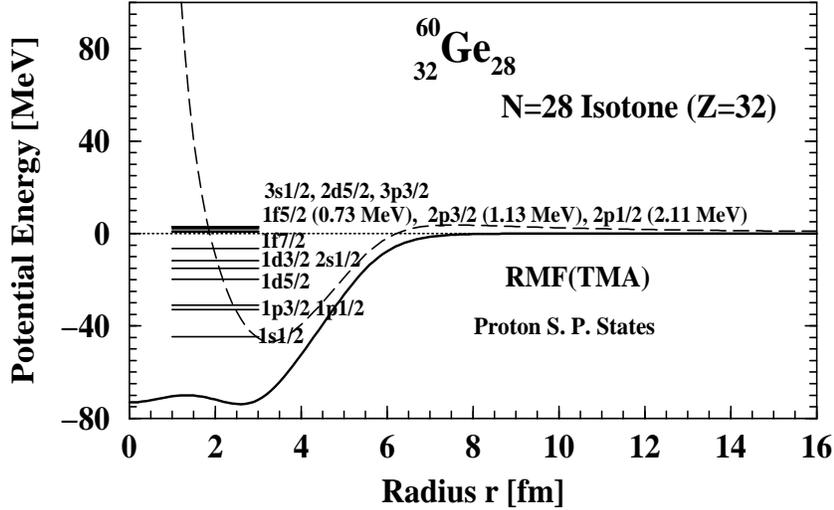,width=11cm,height=7cm}} \caption{The
RMF potential energy for the unbound nucleus $^{60}_{32}$Ge$_{28}$
as a function of radius is shown by the solid line. The long dashed
line represents the sum of RMF potential energy and the centrifugal
barrier energy for the proton resonant state 1f$_{5/2}$. It also
shows the energy spectrum of some proton single particle states
including the important resonant states 1f$_{5/2}$, 2p$_{3/2}$ and
2p$_{1/2}$ at 0.73, 1.13 and 2.11 MeV, respectively.}
\end{figure}

The predicted two-proton emitter nucleus $^{60}_{32}$Ge$_{28}$ is
found to have slightly deformed shape with $\beta_{2m}$ = 0.13 and
has been described here in an approximate manner within the
framework of spherical RMF approach as in the case of nucleus
$^{42}_{24}$Cr$_{18}$. The calculated two-proton separation energy
for the nucleus $^{60}_{32}$Ge$_{28}$ is found to be negative,
though very close to the zero energy. Thus, theoretically this
nucleus appears to be unbound against the two-proton decay. On the
other hand, through the measured value of the two-proton separation
energy for this nucleus is also found to be close to zero, it has
positive sign but with large error bars.

In the case of proton rich $^{60}_{32}$Ge$_{28}$ nucleus the proton
resonant states are found to be 1f$_{5/2}$, 2p$_{3/2}$ and
2p$_{1/2}$ as has been shown in Fig. 11. The figure displays
together the RMF potential and the spectrum for the bound proton
single particle states. We have also depicted in Fig. 11 the total
mean-field potential for the resonant $1f_{5/2}$ state, obtained by
adding the centrifugal potential energy. It may be emphasized that
amongst the many proton single particle states in the continuum only
the resonant state has sizable overlap with the bound single
particle states. Thus the resonant states 1f$_{5/2}$, 2p$_{3/2}$ and
2p$_{1/2}$  make significant contribution to pairing energy.

\begin{figure}[th]
\centerline{\psfig{file=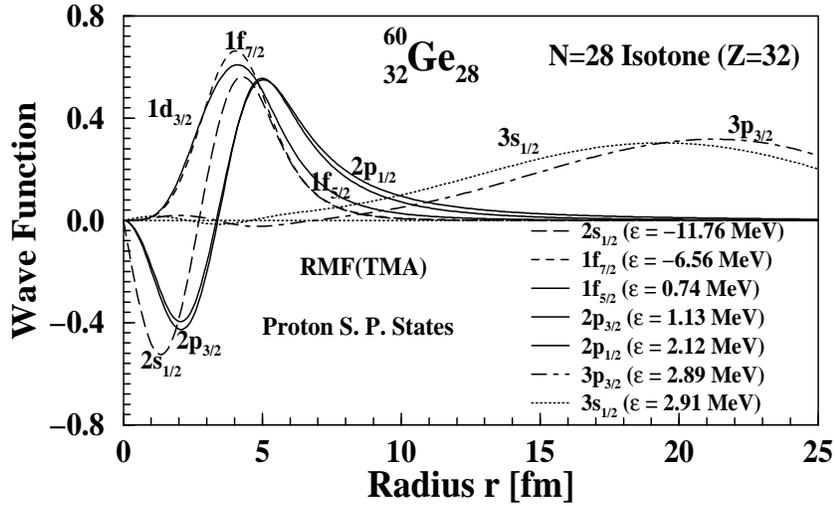,width=11.0cm,height=7cm}}
\caption{Radial wave functions of a few representative proton single
particle states with energy close to the Fermi surface, the proton
Fermi energy being $\lambda_p\, =\,0.64$ MeV, for the nucleus
$^{60}_{32}$Ge$_{28}$. In addition to the bound $2s_{1/2}$ and
$1f_{7/2}$ states, figure shows the resonant states 1f$_{5/2}$,
2p$_{3/2}$ and 2p$_{1/2}$ at 0.74, 1.13 and 2.11 MeV, respectively.
The resonant states similar to the bound states are mostly confined
within the potential region.}
\end{figure}

This can be inferred from Fig. 12 wherein we have displayed the
radial wave functions of some of the proton single particle states
close to the Fermi surface, the proton Fermi energy being
$\lambda_p\, =\,0.64$ MeV. These include the bound  states
$2s_{1/2}$ and $1f_{7/2}$,  and the continuum states $3p_{3/2}$ and
$3s_{1/2}$, in addition to the resonant 1f$_{5/2}$, 2p$_{3/2}$ and
2p$_{1/2}$ single particle states. The wave functions for the
1f$_{5/2}$, 2p$_{3/2}$ and 2p$_{1/2}$ states plotted in Fig. 12 are
clearly seen to be confined within a radial range of about 7 fm and
have a small decaying component outside this region, characterizing
the resonant states. In contrast, the main part of the wave function
for the non-resonant states, e.g. 3s$_{1/2}$ and 3p$_{3/2}$, is seen
to be spread over outside the potential region, though a small part
is also contained inside the potential range. This type of states
thus has a poorer overlap with the bound states near the Fermi
surface leading to small value for the pairing gap $\Delta_{j}$.

\begin{figure}[th]
\centerline{\psfig{file=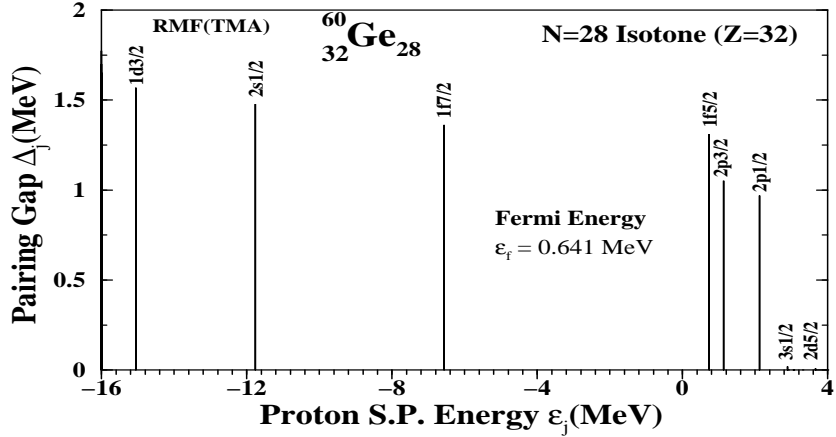,width=11cm,height=6cm}}
\caption{Pairing gap energy $\Delta_{j}$ of proton single particle
states with energy close to the Fermi surface for the nucleus
$^{60}_{32}$Ge$_{28}$.}
\end{figure}

This is clearly evident from Fig. 13 which shows the calculated
single particle pairing gap energy $\Delta_j$  for some of the
proton states in the nucleus $^{60}_{32}$Ge$_{28}$. One observes
indeed in Fig. 13 that the gap energies for the resonant $1f_{5/2}$,
2p$_{3/2}$ and 2p$_{1/2}$ states have values close to $1$ MeV which
is quantitatively similar to that of bound states $1f_{7/2}$ and
$2s_{1/2}$. Also, Fig. 13 shows that the pairing gap value for the
non-resonant states $2d_{5/2}$ and $3s_{1/2}$ lying in the continuum
is negligibly small.

The explanation of this nucleus acquiring long life time despite
lying close to the two-proton drip-line with negative two-proton
separation energy can be given by considering the neighboring
isotones like $^{56}_{28}$Ni$_{28}$, $^{58}_{30}$Zn$_{28}$ in a
manner similar to the case of nucleus $^{42}_{24}$Cr$_{18}$
described earlier.

In order to show the effect of addition of protons to a fixed number
of neutrons which is equivalent to highlighting the changes as one
moves towards the proton drip-line and beyond, we have shown in Fig.
14, the variation in radial density distributions for the isotones
corresponding to N = 20 covering the Z range from 20 to 28.

\begin{figure}[th]
\centerline{\psfig{file=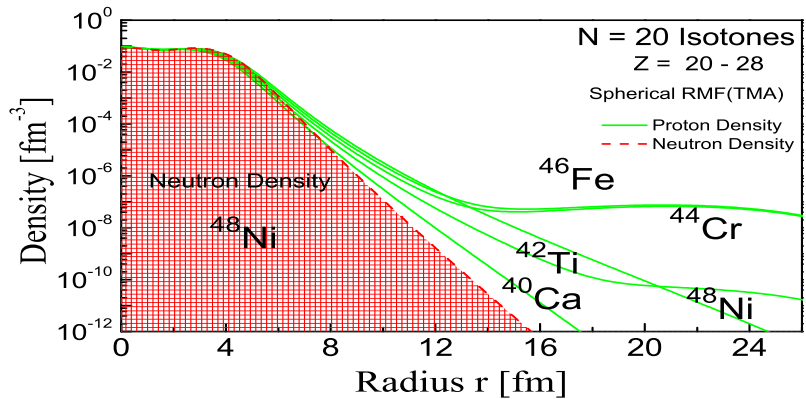,width=11cm,height=6cm}}
\caption{Results for the radial density distributions obtained in
spherical RMF approach using TMA force parameters for N = 20
isotones with proton number Z ranging from 20 to 28. The hatched
area shows the neutron density distributions for the nucleus
$^{48}_{28}$Ni$_{20}$ corresponding to N = 20. The proton density
distributions for different isotones have been shown by solid
lines.}
\end{figure}

It is evidently seen that the proton density has a long extended
tail indicating loosely bound protons. Further increase in number of
protons make the system unbound and in some cases such as
$^{42}_{24}$Cr$_{18}$, $^{48}_{28}$Ni$_{20}$ and
$^{60}_{32}$Ge$_{28}$ discussed here, we have the phenomena of two
proton radioactivity.

\section{Summary}
\label{sect:summary} To, summarize, we have employed deformed
relativistic mean-field plus state dependent BCS (RMF+BCS) approach
to study the ground state properties of proton rich nuclei in the
region 20 $\leq$ Z $\leq$ 40. We have found that the potential
barrier provided by the Coulomb interaction and that due to
centrifugal force may cause a long delay in the decay of some of the
nuclei with small negative proton separation energy. This may cause
the existence of proton rich nuclei beyond the proton drip line.
Nuclei  $^{38}$Ti, $^{42}$Cr, $^{45}$Fe, $^{48}$Ni, $^{55}$Zn,
$^{60}$Ge, $^{63,64}$Se, $^{68}$Kr, $^{72}$Sr and $^{76}$Zr are
expected to be the possible candidates for exhibiting two-proton
radioactivity in the region 20 $\leq$ Z $\leq$ 40. The calculated
two-proton separation energy and other relevant ground state
properties are found to be in good agreement with the available
experimental data. This demonstrates the validity and usefulness of
the deformed RMF+BCS approach for the description of proton rich
nuclei near the drip lines.

The above mentioned results of our investigations for the two-proton
radioactivity are expected to provide additional impetus for more
experimental studies to verify the potential candidates predicted
above for the two-proton radioactivity in the near future.

\section*{Acknowledgments}

Support through a grant (SR/S2/HEP-01/2004) by the Department of
Science and Technology (DST), India, is acknowledged. The authors
are indebted to Dr. L. S. Geng, RCNP, Osaka University, Osaka,
Japan, for valuable correspondence.

\end{document}